\documentclass[
superscriptaddress,
amsmath,amssymb,
aps,
twocolumn,
floatfix,
]{revtex4-2}

\usepackage{graphicx}
\usepackage{dcolumn}
\usepackage{bm}
\usepackage{xcolor}
\usepackage[normalem]{ulem}
\usepackage{amsmath}
\usepackage[english]{babel}
\usepackage{braket}
\usepackage{siunitx}
\usepackage[T1]{fontenc}
\usepackage{float}
\usepackage{chngcntr}

\usepackage{ifthen}
\usepackage{amssymb}
\usepackage{xcolor}

\newboolean{showcomments}
\setboolean{showcomments}{true}

\makeatletter
\newcommand{\mynote}[3]{%
  \ifthenelse{\boolean{showcomments}}{%
   \fbox{\bfseries\sffamily\scriptsize#1}%
   {\small$\blacktriangleright$\textsf{\emph{\color{#3}{#2}}}$\blacktriangleleft$}}%
  {%
   \@bsphack
   \@esphack
  }%
}
\makeatother

\definecolor{asparagus}{rgb}{0.53, 0.66, 0.42}

\newcommand{\MHz}{~$\textrm{MHz}$}

\newcommand{\GHz}{~$\textrm{GHz}$}
\newcommand{\dBm}{~$\textrm{dBm}$}
\newcommand{\dB}{~$\textrm{dB}$}
\newcommand{\K}{~\SI{}{\kelvin}}
\newcommand{\uA}{~\SI{}{\micro\ampere}}
\newcommand{\pH}{~$\textrm{pH}$}
\newcommand{\pF}{~$\textrm{pF}$}
\newcommand{\mK}{~$\textrm{mK}$}

\newcommand{\um}{~\SI{}{\micro\meter}}
\newcommand{\cm}{~\SI{}{\centi\meter}}
\newcommand{\mm}{~\SI{}{\milli\meter}}
\newcommand{\us}{~\SI{}{\micro\second}}
\newcommand{\ns}{~\SI{}{\nano\second}}

\usepackage{tabularx}
\usepackage{graphicx}
\usepackage{dcolumn}
\usepackage{bm}

\begin{document}

\preprint{AIP/123-QED}

\title{Simple, High Saturation Power, Quantum-limited, RF SQUID Array-based Josephson Parametric Amplifiers}

\author{Ryan Kaufman}
\email{rrk26@pitt.edu}
\affiliation{Department of Physics and Astronomy, University of Pittsburgh, Pittsburgh, PA}
\author{Chenxu Liu}
\affiliation{Physical and Computational Sciences, Pacific Northwest National Laboratory, Richland, WA}
\author{Katarina Cicak}
\affiliation{National Institute of Standards and Technology, Boulder, CO}
\author{Boris Mesits}
\affiliation{Department of Physics and Astronomy, University of Pittsburgh, Pittsburgh, PA}
\author{Mingkang Xia}
\affiliation{Department of Physics and Astronomy, University of Pittsburgh, Pittsburgh, PA}
\author{Chao Zhou}
\affiliation{Department of Applied Physics, Yale University, New Haven, CT}
\author{Maria Nowicki}
\affiliation{Department of Physics and Astronomy, University of Pittsburgh, Pittsburgh, PA}
\author{Jos\'e Aumentado}
\affiliation{National Institute of Standards and Technology, Boulder, CO}
\author{David Pekker}
\affiliation{Department of Physics and Astronomy, University of Pittsburgh, Pittsburgh, PA}
\author{Michael Hatridge}
\affiliation{Department of Physics and Astronomy, University of Pittsburgh, Pittsburgh, PA}

\date{\today}

\begin{abstract}
High-fidelity quantum non-demolition qubit measurement is critical to error correction and rapid qubit feedback in large-scale quantum computing. High-fidelity readout requires passing a short and strong pulse through the qubit's readout resonator, which is then processed by a sufficiently high bandwidth, high saturation power, and quantum-limited amplifier.  We have developed a design pipeline that combines time-domain simulation of the un-truncated device Hamiltonian, fabrication constraints, and maximization of saturation power.
We have realized an amplifier based on a modified NIST tri-layer Nb fabrication suite which utilizes an array of 25 radio frequency Superconducting QUantum Interference Devices (rf SQUIDS) embedded within a low-Q resonator powered by a high-power voltage pump delivered via a diplexer on the signal port. We show that, despite the intensity of the pump, the device is quantum-efficient and capable of high-fidelity measurement limited by state transitions in the transmon. We present experimental data demonstrating up to $-91.2\dBm{}$ input saturation power with $20\dB{}$ gain, up to $28\MHz{}$ instantaneous bandwidth, and phase-preserving qubit measurements with $62\%$ quantum efficiency.
\end{abstract}
\maketitle

\section{Introduction}

Superconducting quantum computers are rapidly scaling to thousands of physical qubits~\cite{IBMRoadmap}; more,  high-fidelity qubit readout is a requirement for both Noisy Intermediate Scale (NISQ)~\cite{Preskill} and error-corrected operation of these machines.  To avoid further inflation of machine size via the use of ancilla qubits for error correction and other algorithms involving mid-circuit measurements, we require the readout to be Quantum Non-Demolition (QND), which leaves the qubit in the state indicated by the measurement outcome~\cite{Nielsen2010}.
Moreover, dispersive superconducting qubit readout is limited by as yet imperfectly understood non-QND effects for strong readout drives~\cite{Sank2016, Shillito2022, Khezri2022, Cohen2023}, and so one cannot increase the power of the measurement indefinitely to increase measurement fidelity. Further, due to the qubit's finite $T_1$, one must measure for a time much smaller than $T_1$ in order to attain high fidelity. Finally, we need a faithfully linear amplifier to amplify the short, few-photon quantum signal to be unaffected by the added noise of standard 4K commercial amplifiers in the upper stage of the readout chain while adding the minimum amount of noise allowed by quantum mechanics~\cite{cavesquantumlimit, Clerk2010}. 

Generally, we require these high-fidelity, QND measurements for each element of a quantum computer. Frequency multiplexing is a natural solution, allowing many qubit readout resonators to be measured from a single feed line as long as their resonant frequencies are sufficiently spaced. However, this spacing has to be larger than a few linewidths of the resonators to avoid crosstalk, necessitating a large bandwidth~\cite{Mutus2014, Roy2015, White2023, Kaufman2023} of the first amplifier in the chain. The presence of many signals at the same time also requires that the first amplifier be highly linear, both avoiding compression and excess intermodulation distortion~\cite{White2023, Kaufman2023}. 

The Josephson Parametric Amplifier (JPA) is one type of first-stage amplifier that shows promise for its ease of manufacture. The JPA's megahertz-scale bandwidth typically falls well short of Josephson Travelling Wave Parametric Amplifiers (JTWPAs). However, the difficulty of manufacturing the thousands of junctions required for JTWPAs, their intrinsic loss, and noise-rise near saturation~\cite{Remm2023}, motivates the study of how we can improve the JPA to create a small, easy-to-fabricate, quantum-limited amplifier. The primary shortcoming of simple JPAs with a single junction or loop is that they have very limited saturation power, typically less than $-110\dBm{}$~\cite{Yamamoto2008, Mutus2014, Zhou2014, Roy2015, Frattini2018}.  Depending on the choice of qubit parameters, this power may not be enough to amplify more than one or two qubits without distortion from saturation effects within the amplifier~\cite{White2023}.
Diluting the nonlinearity of single junction/loop devices using a chain of DC SQUIDs, rf SQUIDs, and SNAILs has been explored both theoretically and experimentally, yielding substantially higher saturation power~\cite{Kochetov2015, Frattini2018, Sivak2019, White2023, Kaufman2023}.

In this work, we design, fabricate, and characterize an extremely quantum-efficient and high-saturation power amplifier using a simple, single-mode device. The key component of our design process is numerical simulations of the un-truncated device Hamiltonian to fully understand and optimize the device within the constraints of fabrication. The key tuning parameters that we optimized are the device quality factor/impedance and the rf SQUID shunting ratio. The resulting simple design consists of merely two capacitors and an rf SQUID array using 25 junctions. We experimentally realized this device in a modified NIST tri-layer process.  We demonstrate input referred saturation powers averaging $-94.2\dBm{}$ ($\pm~1.4\dBm{}$) with $20\dB{}$ of power gain with extremes approaching $-91\dBm{}$ input saturation power and $28\MHz{}$ instantaneous bandwidth using only 25 $5.5\uA{}$ junctions. Additionally, we show that our amplifiers have high quantum efficiency (62\% referred to the plane of the qubit), allowing fidelity measurements that are only limited by the qubit's ability to tolerate photons in the readout resonator, not the amplifier's saturation power.

\section{Amplifier Circuit Design and Fabrication}
In this section, we present the device simulation and design within the bounds of fabrication. We begin by discussing the shortcomings of previous simulation efforts that truncate the nonlinearity in the device Hamiltonian. Following this, we discuss why we choose rf SQUIDs as our nonlinear elements. We also elaborate on our simulation method, which does not require truncation, and optimize the design parameters within our circuit. Lastly, we translate these optimization results into circuit elements with an added impedance transformation to control device quality factor.

The problem with single-junction or single-loop JPA's is that even if their bandwidth is broadened to amplify the readout signals of multiple resonators, they struggle to reach the input saturation powers required to simultaneously amplify multiple readout signals at the $20\dB{}$ gain required to overwhelm the added noise at the next stage of amplification at 4K~\cite{Mutus2014, Roy2015}. Measurements of this effect are shown directly in~\cite{White2023} in readout on a superconducting processor. This problem of JPAs lacking power handling can be traced to the current through and phase difference across the Josephson junction at large input signals, which other works have focused on in some other similar amplifier designs~\cite{Kochetov2015, Frattini2018, Sivak2019}. 

These past works depend crucially on expansions of the junction nonlinearity around a particular bias point to find an analytical expression to guide design. While helpful to guide the designer's intuition at small signal strengths, truncations to the junction Hamiltonian have been shown to obfuscate the high signal power behavior of Josephson junction-based devices~\cite{Liu2020, Frattini2018, Sivak2019}, sometimes requiring extensions up to seven or eight orders to capture the nonlinear behavior of the junctions near amplifier saturation~\cite{Liu2020}. Because one of our chief goals is to increase amplifier saturation power, we instead simulate the device at large signal powers by finding a periodic steady-state solution of the device's classical equation of motion in the time domain. This method is not without its drawbacks.  Notably, it is more computationally intensive than methods such as harmonic balance and does not give any analytical understanding of the design. However, crucially, time domain simulations also do not require any simplification or truncation of the sinusoidal nonlinearity of the junction, and so are a far more accurate guide for the design of high-saturation power amplifiers.

\begin{figure}
\includegraphics[width=3.3in]{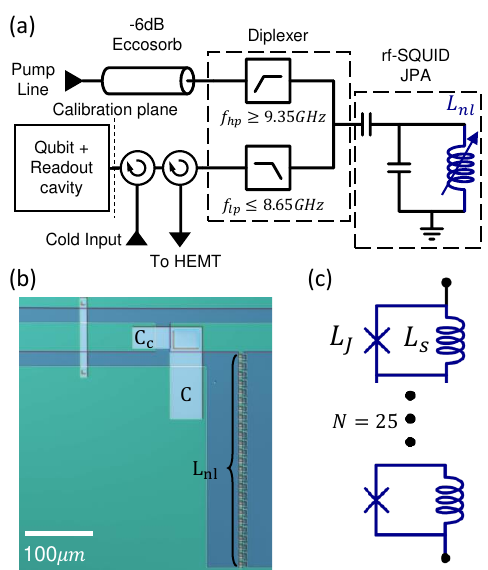}
\caption{\label{fig:circuitFig} (a) The configuration of the measurement circuit in the mixing chamber of a dilution refrigerator. A full diagram is given in Appendix~\ref{"exptAppendix"}. The input power and quantum efficiency of the device are both calibrated from the plane shown at the readout resonator. (b) Device micrograph, showing the coupling capacitor $C_c$ and the main capacitor $C$ that are used to control the quality factor and resonant frequency, respectively. The 25 rf-SQUID long array is shown at right, with approximately a $250\um{}$ span to the ground plane (c) Schematic of the nonlinear inductive element (labeled $L_{nl}$ in (a)) of the amplifier.
}
\end{figure}

To guide our design, we start with the same strategy as other work with DC SQUIDs~\cite{Kochetov2015} and Superconducting Nonlinear Asymmetric Inductive eLements (SNAILs)~\cite{Frattini2017, Frattini2018}.
If the oscillating phase of the signal increases to nearly $2\pi$, higher-order nonlinearities in the sinusoidal potential of the junction begin to become comparable to the third-order nonlinearity required for gain and disrupt amplification. This has been shown previously in similar studies of the same type of behavior in Josephson ring modulators~\cite{Liu2017, Liu2020} as well as Josephson bifurcation amplifiers~\cite{Kochetov2015}. It has also been shown with fourth-order Kerr nonlinearity in SNAIL parametric amplifiers~\cite{Frattini2017, Frattini2018, Sivak2019, Sivak2020}. All of these devices utilize a nonlinear shunt across the junction, formed by Josephson junctions of equal or higher critical currents.   
For high-inductance versions of these loops, an array of JJs is desirable for producing rather linear arrays with small geometric size.  However, for large critical currents/small inductances, such as we use in amplifiers, it becomes quite feasible and desirable to use the geometric inductance of a short superconducting lead as the shunt, which has the additional advantage of eliminating concerns about array modes of the shunt itself \cite {Masluk2012, Frattini2018}.



\begin{figure}[!ht]
\includegraphics[width=3.37in]{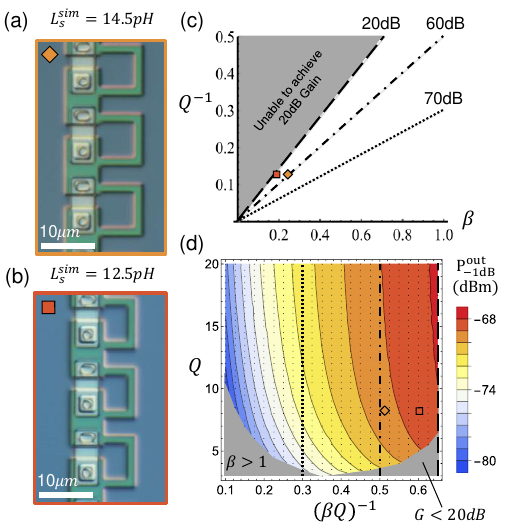}
\caption{
(a) Micrograph of three out of the 25 rf-SQUID loops in device A. (b) Micrograph of three out of the 25 rf-SQUID loops in device B. (c) The design space of a single-mode degenerate JPA as a function of shunt inductance (for a given junction inductance) and quality factor Q of the JPA mode in steady state at $\omega$ when pumped at $2\omega$. The junction inductance $L_J \approx 60\pH{}$ was chosen for convenience given by a preexisting reliable fabrication process. When combined with the choice of a reasonably low quality factor $Q$, this determines the choice of $\beta$, shown altogether as an orange diamond in the design space. An additional device B with a lower simulated shunt inductance is shown as a red square. (d) The predicted output $P_{1dB}$ of devices in a subsection of the design space. The dependence on the slopes in (c) is emphasized by dotted/dashed lines, which in this coordinate space are vertical.}
\label{fig:theoryFig}
\end{figure}

This device we refer to as an rf SQUID, though we acknowledge that is method of operation and biasing is rather different than the historical devices given this name  \cite{Clarke2005}.  We then carefully consider how best to optimize the circuit to maximize saturation power. To complement the dilution of the current through the junction by the linear inductive shunt, the phase is distributed by even division over an N-element series array similar to past work in Refs \cite{Kochetov2015} and \cite{ Frattini2018}. The full circuit with all external circuit elements is shown in Fig.~\ref{fig:circuitFig}(a). The result is similar to designs proposed in~\cite{Naaman2019} but simpler, with only one array instead of two in parallel, albeit without an easy means of delivering flux on-chip over the entire array.

To simulate the device, we begin with the circuit diagram in Fig.~\ref{appfig:TheoryDesign}(a), with a single nonlinear mode formed from an array of $N$ rf SQUIDs of shunt inductance $L_s$ in parallel with a Josephson junction with inductance $L_J$ resonated by a capacitor $C$, and coupled to the external environment of characteristic impedance $Z_0$ around resonance. Assuming that the external magnetic fluxes through the rf SQUID,noted as $\phi_\text{ext}$, are the same, the equation of motion (EOM) can be expressed as
\begin{equation}
\begin{split}
    \partial_t^2 \phi(t) + \gamma \partial_t \phi(t) +\omega_L^2 \phi(t) +\omega_J^2 \sin \left[ \frac{\phi(t)+ \phi_\text{ext}}{N} \right] \\ = 2 \gamma \partial_t \phi_{in}(t),
    \label{Eq:EOM_pre}
\end{split}
\end{equation}

\noindent where the phase variable is defined by $\phi(t) = \int V(t) dt$ (see Fig.~\ref{appfig:TheoryDesign}a), $\gamma = (C Z_0)^{-1}$ is the external decay rate, $\omega_L^2 = 1/(C L_s \cdot N)$ and $\omega_J^2 = 1/(C L_J)$ are two frequency constants.

Using this simulation method, we find that controlling the dimensionless ratio of shunt to Josephson inductance $L_J$, $\beta = L_s/L_J$ of each rf SQUID in the array and the quality factor $Q$ of the resonator is critical to maximizing the saturation power of the rf SQUID while maintaining its ability to be pumped to produce gain. We find that the device should have approximately the same maximum achievable gain for constant $(\beta Q)^{-1}$, shown in Fig.~\ref{fig:theoryFig}(c) by different dashed/dash-dot/dotted lines through the design space. This is qualitatively similar to discussions about the $pQ$ product in SNAIL and JPC parametric amplifiers, with $Q$ having the same meaning as here, and $p$ being a ratio of SNAIL inductance to the total inductance of the mode. Much like in our rf SQUID case, there a higher $pQ$, analogous to a lower $(\beta Q)^{-1}$, is "safer" in the sense that it ensures the amplifiers will achieve 20dB of gain~\cite{Bergeal2010, Frattini2018}.

From our numerical calculations, we find that the design parameters Q and $\beta$ roughly outline a region where the amplifier will fail to achieve $20\dB{}$ gain if the rf SQUID is shunted with too small an inductor or the quality factor is too low, as shown by the dashed line determined by a slope $(\beta Q)^{-1} \approx 0.65$. For every point in the design space, we choose a flux bias point that maximizes the third-order nonlinearity in the rf-SQUID. However, we also find that variations in the saturation power with respect to flux bias are small in theory (see Fig.~\ref{appfig:TheoryConfigFig}(b)). Our principal result is that we also find that the maximum saturation powers of the device occur on the boundary outlined by the same dashed line through the design space in Fig.~\ref{fig:theoryFig} (c) and (d), and can increase the saturation power in a 25-element long array by over $10\dB{}$ compared to the lowest values. A higher inductive shunt produces a more nonlinear device with lower saturation power for a given Q, but more tunability in its inductance with respect to flux bias. For mode impedances that are controllably low (such that stray parallel capacitance and series inductances are small compared to the LC resonator that makes up the mode) at $5\GHz{}$, this sensitivity corresponds to requiring single micron-scale control over the exact dimensions of the inductive shunt of each rf SQUID in the array. Fig.~\ref{fig:theoryFig}(d) shows the design space of the amplifier assuming an $N=25$ rf-SQUID array in an embedding resonator with $\omega_0/2\pi \approx 6\GHz{}$.

Based on this understanding, we next scale the circuit by arraying N rf SQUIDs in series. A larger array lowers the nonlinearity by diluting the phase across each rf SQUID, raising the amplifier's saturation power, but also making it harder to pump. Scaling the array by holding total array inductance constant and decreasing both the individual junction and shunt inductances as the array number increases would increase the saturation flux linearly. This yields quadratic saturation power scaling if one could maintain a constant mode impedance by scaling the junction critical current up and the shunt inductance down indefinitely. However, scaling the junction size up decreases its inductance, and since the designer has to maintain the same shunting ratio, the shunt inductance becomes difficult to control with optical lithography below about $10\pH{}$ if the junction size is a few microns. Additionally, the junction side of the rf-SQUID has a stray inductance of a few$\pH{}$, which can become comparable to the shunt as the scaling is increased. In combination, these effects put a lower bound on the achievable shunt inductance, and therefore an upper bound on the critical current of the junction. With this restriction in mind, we choose to use 25 rf SQUIDs that only require a comfortably low inductance $12-15\pH{}$ shunt inductor and a $5.5\uA{}$ junction as the basic ingredient for our array.



Our fabricated devices, shown in Fig.~\ref{fig:theoryFig}(a) and (b),  are fabricated in a modified niobium trilayer process at the National Institute for Standards and Technology's Boulder Microfabrication Facility (layer specifications given in Appendix~\ref{"sec:fab_layers"}), where the high yield and consistency  of the junctions allows for excellent device yield in even very long junction arrays~\cite{Niemeyer1985yield}. Even so, spatial nonuniformity in the flux biasing in the array (discussed in Appendix~\ref{sec:flux uniformity} and \ref{"sec:flux offsets"}) presents an issue with verifying device parameters in the experiment. Accordingly, we discuss only the nominal designed values here, with the understanding that these parameters can reasonably vary by 10 percent or so in the real device. We use a shunting ratio of $\beta=0.25$ for device A, 0.21 for device B, and a junction inductance of $60~\textrm{pH}$.

In total, the rf SQUID array forms about $290\pH{}$ of total effective inductance at zero DC-bias, and between $340$ to $390\pH{}$ at the flux bias points which we expect to achieve $20\dB{}$ gain. Provided this inductance is paired with a resonant capacitor that produces $\omega_0/2\pi \approx 6\GHz{}$ we arrive at a capacitor value of $2\pF{}$.  If this mode were directly connected to a $50 \Omega$ transmission line, we would expect a quality factor $Q\approx4$. However, as mentioned above, we have to alter the quality factor due to the increased inductance of the array compared to the single rf-SQUID. To increase the quality factor well into the region where we expect to achieve $20\dB{}$ of gain (see Fig.~\ref{fig:theoryFig}(d)), we add a coupling capacitor $C_c\approx0.26\pF{}$ to increase the effective environmental impedance on resonance to $\approx100 \Omega$ as seen by the JPA resonator, and increase the quality factor to approximately $Q=10$, with some added variance over bias flux because of the coupling capacitor's frequency-dependent impedance. The effect of this variation with respect to flux and its consequences for the tunability of the amplifier's gain is discussed in Appendix~\ref{"sec:gain tuning"}.

We measure the packaged JPA chip using a typical qubit readout amplification chain, with a few additions, such as eccosorb filters, a diplexer, and a modified pump line, that are detailed in Appendix~\ref{"sec:lines"}.

\section{Results}

\begin{figure}[h]
\includegraphics{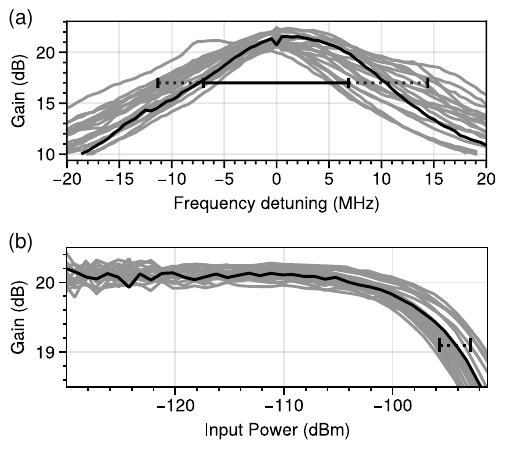}
\caption{(a) Gain profiles with varying pump-detuning, with bandwidth deviations shown. (b) Saturation power of the $N = 25$, $14.5\pH{}$ shunted rf-SQUID array amplifier at multiple bias points. In both plots, the black lines are particular bias points that have similar values to the mean. The means and standard deviations are calculated from the full dataset shown in Fig.~\ref{appfig:satPwrVsBias}.
}\label{fig:gainSatFig}
\end{figure}

Cooled to $15\mK{}$ in a dilution refrigerator, and powered with a strong pump at $2\omega$ the 25-long rf-SQUID array shown in Fig.~\ref{fig:theoryFig}(a) and in the larger device in Fig.~\ref{fig:circuitFig}(b), is capable of generating a three-wave gain of more than $20\dB{}$ over a range of bias points, shown in aggregate in Fig.~\ref{fig:gainSatFig}(a) with more detail in Fig.~\ref{appfig:satPwrVsBias}(b). To be sure we account for any loss introduced by the amplifier, we take the maximum return loss over all bias fluxes at each frequency as the baseline for gain. After using this pessimistic estimate to tune to $20\dB{}$ of gain at each bias point, we measure input signal compression power up to a maximum of $-91.5\dBm{}$, which is among the highest values reported in the literature for a Josephson junction-based resonant parametric amplifier. This value is in good company with the extreme values of~\cite{Frattini2018, Sivak2019, Sivak2020}, and the average values in other rf-SQUID based devices~\cite{White2023, Kaufman2023}. Notably, devices of this type achieve similar saturation powers even to JTWPAs that contain almost one hundred times more junctions~\cite{Macklin2015}. 

Power is calibrated at $5.7\GHz{}$ by measuring resonator photon number-induced shifts of the qubit's transition frequency ($\omega_{01}/2\pi$) combined with separate measurements of the qubit-cavity dispersive shift, shown in Appendix~\ref{"sec:sat pwr cal"} along the lines of~\cite{Schuster2007}.

Average input saturation power values are $(-94.2\pm1.4)\dBm{}$, with an average instantaneous bandwidth of $(20\pm6)\MHz{}$. As shown in Fig.~\ref{fig:gainSatFig}(b) with more detail in Fig.~\ref{appfig:satPwrVsBias}, the gain varies somewhat in profile with respect to pump detuning at any given bias point, which we attribute to ripples in the line impedance. Further, because the pessimistic baseline we have chosen for gain is not necessarily flat with respect to frequency and is combined from all bias points, some of the gain profiles are asymmetric with respect to $\omega_p/2$. If we take a less pessimistic baseline, only normalizing to one flux bias point at a time, they appear symmetric as expected. Some variation in the shape of the loss is expected with the device's low quality factor because the device responds to changes in the external environment over its linewidth of approximately $300-500\MHz{}$, with variations over flux bias. This significantly impacts the pumped operation of the device, because the characteristic ripples in the external environment occur on the same $100\MHz{}$ scale. Consequently, the gain profile should take on a modified profile depending on the local slope of the external environment at that particular signal frequency. In addition to the shapes shown, it is possible to see double-peaked gain profiles similar to~\cite{Roy2015}. We fall short of the expected $\sqrt{GB}$ gain-bandwidth product for many bias points partly because of this issue of variable port impedance. There is also significant uncertainty in the fit for the device linewidth due to the same impedance ripples, so it is possible the device linewidth may be lower than designed.

The input-referred saturation power of the amplifier depends slightly on bias settings and certain parameters that vary on the $1\dB{}$ level in the external environment,  such as the loss between the resonator cavity and the amplifier chip. Within these variations, the data shows reasonable agreement with the theoretically predicted value from periodic-steady-state analysis of the device equation of motion, shown in Fig.~\ref{appfig:TheoryConfigFig}.
The high saturation power of the rf-SQUID JPA that we demonstrate is critical for enabling many-qubit readout experiments. However, despite the diplexer's rejection, the pump power (approximately -25 to -30dBm, based on room temperature loss calibrations) required and the lack of attenuation on the pump line might raise concerns over the thermal photon occupation in the readout resonator and the corresponding limits imposed on $T_2$ from thermal photons in the readout resonator dephasing the qubit. 



\begin{table}[!ht]
\begin{tabularx}{\linewidth}
{
| >{\raggedright\arraybackslash} l |
| >{\centering\arraybackslash} X 
| >{\centering\arraybackslash} X 
| >{\centering\arraybackslash} X|}
\hline
\textrm{Amplifier State}&
\textrm{$T_1$ ($\mu s$)}&
\textrm{$T_{2,R}$ ($\mu s$)}&
\textrm{$T_{2,E}$ ($\mu s$)}\\
\hline
Replaced with short & 81 & 8 & 20\\
\hline
Off & 90 & 13 & 26\\
\hline
Detuned -500MHz & 93 & 11 & 18\\
\hline
On (25dB Gain) & 91 & 6.4 & 9.1\\
\hline
\end{tabularx}

\caption{Qubit coherence versus different states of the amplifier. The run in which the amplifier was disconnected was measured separately from the other three. The $T_2$ coherence differences with respect to the amplifier's state suggest that the reverse isolation between the amplifier and the qubit may not be sufficient.}\label{tab:coherence}
\end{table}

To address this concern, we measure qubit coherence with varying pump conditions in Table~\ref{tab:coherence}. We speculate that the presence of the high-power pump line is not meaningfully contributing to decoherence at the $8\us{}$ level, otherwise we would have expected that the qubit coherence would have improved in the absence of the amplifier and diplexer combination in row 1. It must be noted, however, that this measurement was taken on a different cooldown than the others, and qubits can have significant variation in coherence from cooldown to cooldown. Avoiding dephasing is crucial to the design of high-saturation JPAs, allowing the pump inefficiency to be tolerable in terms of qubit coherence albeit inconvenient for fridge design, requiring specially engineered lines with lower attenuation in order to deliver high power pumps without excess heating. However, we observe the $T_{2R}$ time of the qubit dip distinctly when the amplifier is turned on, indicating that the amplifier introduces an extra source of dephasing onto the qubit from the gain alone. This effect is often observed in high-efficiency measurement setups~\cite{Lecocq2017, Lecocq2021}, and can be the result of insufficient isolation between the JPA and the readout resonator. To combat this, one can use additional circulators for isolation at the cost of a modest loss in quantum efficiency~\cite{Kaufman2023}. 

\begin{figure}[!ht]
\includegraphics[width = 3in]{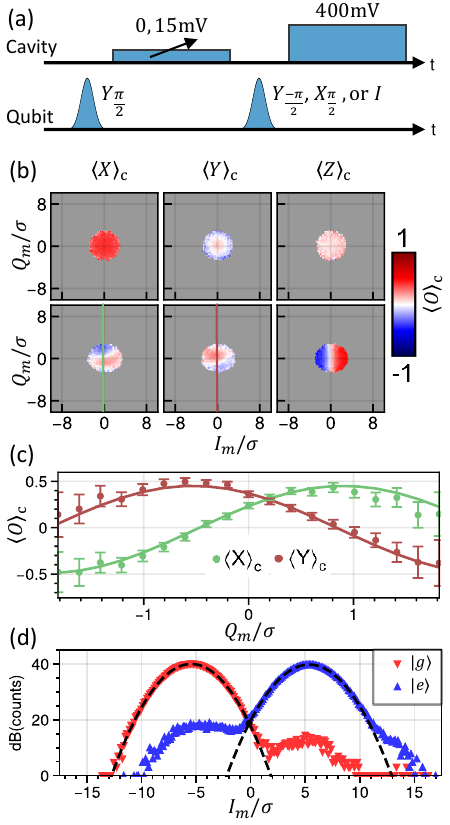}
\caption{
(a) Tomography of variable-strength measurement of a transmon qubit prepared in the $|+X\rangle$ state, with the amplifier biased to $25\dB{}$ of phase-preserving gain. (b) Conditional measurement back-action in the X and Y axes with respect to I and Q, increasing in frequency with measurement strength, is a measure of efficiency, showing sinusoidal oscillation in the perpendicular axis to measurement in the X and Y back-action of the qubit. As measurement strength increases, the contrast decreases as the qubit state collapses to the eigenstates of the measurement more quickly, but the oscillation frequency increases. (c) A line cut at $I_m = 0$ is fit to yield $(62.4\pm1.4)\%$ quantum efficiency, normalized to 1. The data shown are the averages from a Gaussian distribution of samples that varies over $Q_m$, with the variation in samples reflected in the error bars of the fit. (d) Gaussian fitting to an integrated $800\ns{}$ readout pulse to determine $99.3\%$ separation fidelity, integrated from an IQ histogram containing over $10^6$ counts. Red hue indicates counts in the qubit state $\ket{g}$, while blue indicates counts in $\ket{e}$.
}
\label{fig:sciprotocol}
\end{figure}

In order to calibrate the efficiency of the amplification chain from the plane of the qubit, we can measure the back-action that the amplifier introduces on the qubit. We turn the amplifier on at $25\dB{}$ gain, prepare the qubit in the $\ket{+X}$ state, and apply two measurements in sequence according to the pulse diagram in Fig.~\ref{fig:sciprotocol}(a)~\cite{Hatridge2013}.

The back-action of the weak measurement results in a sinusoidal pattern in the $\left<X\right>_c$ and $\left<Y\right>_c$ expectation values in Fig.~\ref{fig:sciprotocol}(b), with linecuts in Fig.~\ref{fig:sciprotocol}(c) with respect to the $ Q_m $ result of the weak measurement. This sinusoidal pattern has frequency set by the measurement strength $\overline{I_m}/\sigma$, but the amplitude is diminished by an exponential dependent on the measurement strength scaled by the quantum efficiency~\cite{Hatridge2013}. Because measurement strength is independently set by the distance between coherent pointer states in the IQ plane, we can fit these oscillations to extract the quantum efficiency. Because the IQ values are dependent on the entire readout chain, the quantum efficiency $ \eta $ that results from the fit represents the efficiency of the entire readout chain, including losses and added noise from the higher-stage amplifiers at $4\K{}$ and room temperature. Therefore, it acts as a safe lower bound for the efficiency of the rf-SQUID JPA itself.

With this method, we find a measurement efficiency of $(0.62~\pm0.014)$ on a scale of 0 to 1, without compensating for state preparation and measurement error. This is among the highest reported quantum efficiency values for phase-preserving amplifiers~\cite{Walter2017, Touzard2019, White2023, Kaufman2023}, however still not as efficient as phase-sensitive nonreciprocal amplification~\cite{Lecocq2021}, serving as motivation to pursue integrating amplification closer to the readout resonator itself. The high quantum efficiency of our device also serves as evidence that the high-power nature of the rf-SQUID JPA does not detract from its efficiency.

The projective state readout fidelity of the qubit suffered from a below optimal $2\chi/\kappa = 0.348$, where $2\chi = 0.348\MHz{}$ is the change in readout resonator frequency between $\ket{g}$ and $\ket{e}$ states of the qubit and $\kappa = 1\MHz{}$ is the readout resonator linewidth. So, despite high readout efficiencies, the $T_1$ decay of the qubit limits the phase-preserving readout fidelity to $99.3\%$ with a $0.8\us{}$ pulse length, as shown in Fig.~\ref{fig:sciprotocol}(d). As a result, despite the high saturation power of the amplifier, we were unable to increase the readout power and shorten the pulse time due to readout-induced state transitions in the qubit. For examples of high-power readout histograms, see Fig.~\ref{appfig:highPWRMsmt}.

Higher state transitions in qubit-cavity systems at high cavity occupation is an oft-observed phenomenon~\cite{Sank2016, Khezri2022, Shillito2022, Cohen2023}. However, it is not entirely without counter. One can design linear filters that take into account quantum noise and $T_1$ decay~\cite{Gambetta2007, Khan2024}. Even with such filters the behavior of transmon qubits at very high readout resonator occupations is generally detrimental to readout performance, and remains a barrier to fast, high-fidelity, QND measurement. Future work to improve the state of the art, including the study of transmon dynamics at high power, will hopefully be enabled by a robust measurement chain that can easily tolerate the measurement powers that drive transmons into behavior resembling chaos~\cite{Cohen2023}.
\section{Conclusion}
To conclude, we have experimentally demonstrated the effectiveness of periodic steady-state simulations for capturing JPA dynamics. Using these tools we have successfully designed, built, and fabricated a high saturation power JPA built on rf-SQUID arrays with enough power handling capability for hundreds of channels of simultaneous measurements, albeit without the bandwidth to demonstrate this capability. Additionally, we have shown that despite the device's need for a high-power pump, it does not reduce qubit coherence beyond what is expected for a single stage of isolation, which can be further circumvented by pulsing the amplifier. In addition, the device maintains quantum efficiency on par with some of the best examples of phase-preserving amplification in the literature. We show high-fidelity measurement limited primarily by readout-activated state transitions in the qubit at high powers, rather than any property of the amplifier itself.  The non-QND nature of strong measurement tones in circuit QED remains an open problem for the field which motivates the pursuit of optimal filtering~\cite{Khan2024} and a greater understanding of the limitations of dispersive readout in circuit QED~\cite{Shillito2022, Cohen2023, Dumas2024}.

Future work includes engineering a robust matching network to broaden the gain response of this amplifier with techniques similar to~\cite{Naaman2022, Kaufman2023}. However, we note that while these techniques outline an elegant solution that traces roots back to filter synthesis, they must also be unified with the saturation power calculations demonstrated in this work which account for the full nonlinearity of the Josephson junction to further assist the designer in anticipating saturation effects. More, we should implement a dedicated pump coupling port that can avoid the inefficiency of the coupling capacitor as a means of delivering the AC driving phase to the junction at the pump frequency. This could effectively integrate the external diplexer on-chip, which can be feasibly implemented with the existing fabrication process, which will greatly simplify the external circuitry required to operate the amplifier and conceivably reduce the loss between the amplifier and qubit to further increase the efficiency of measurements. Additional research is also being done on how to increase the amplifier's power-added efficiency once the pump reaches the resonator node~\cite{Hougland2024}. In all, we demonstrate a simple and highly linear amplifier that can serve as a stepping stone for these innovations and as a workhorse in near-term quantum information experiments, including the study of the QNDnesss of strong qubit readout.

\section*{Acknowledgements}
We are grateful to the Peterson Institute for Nanoscience and Engineering for support in the fabrication of the qubit used for measurement. We thank Chao Zhou and Pinlei Lu for their work and helpful guidance on the control and measurement software. 
This manuscript is
based upon work supported in part by the U.S. Army
Research Office under Grant No. W911NF-15-1-0397 and the National Institute of Standards \& Technology (NIST) Quantum Information Program. Boris Mesits was supported by the NSF GRFP. This material is based upon work supported in part by the U.S. Department of Energy, Office of Science, National Quantum Information Science Research Centers, Co-design Center for Quantum Advantage (C2QA) under contract number DE-SC0012704, (Basic Energy Sciences, PNNL FWP 76274). The Pacific Northwest National Laboratory is operated by Battelle for the U.S. Department of Energy under Contract DE-AC05-76RL01830.
This is a contribution of the National Institute of Standards and Technology, not subject to U.S. copyright. Any identification of commercial equipment, instruments, or materials in this paper is to foster understanding; it does not imply neither the recommendation or endorsement by the National Institute of Standards and Technology, nor does it imply that the materials or equipment identified are necessarily the best available for the purpose.

\bibliography{main}
\pagebreak
\clearpage

\appendix
\begin{widetext}
    \counterwithin{figure}{section}
    \section{Periodic steady-state simulation}\label{"theoryAppendix"}

We analyze the dynamics of the rf-SQUID amplifier using periodic steady-state simulation similar to the methods used to analyze Josephson Ring Modulators in \cite{Liu2020}, starting Eq.~\eqref{Eq:EOM_pre} in the main text. Note that the phase variable $\phi(t)$ has DC components, noted as $\phi_\text{DC}$, and hence $\phi(t) = \varphi(t) + \phi_\text{DC}$. The DC component can be determined by the current relation in each rf-SQUID, 
\begin{equation}
    \frac{\phi_\text{DC}}{L_s \cdot N} + \frac{1}{L_J} \sin \left( \frac{\Delta \phi}{N} \right) = 0,
\end{equation}
where we define $\Delta \phi = \phi_\text{DC} + \phi_\text{ext}$.

By separating the DC and AC components, we rewrite the EOM in Eq.~\eqref{Eq:EOM_pre} as
\begin{equation}
    \partial_t^2 \varphi(t) + \gamma \partial_t^2 \varphi(t) +\omega_0^2 \varphi(t) +\omega_J^2 \left[ \sin \left(\frac{\varphi(t) + \Delta \phi}{N} \right) -\sin\left( \frac{\Delta \phi}{N} \right) - \frac{\varphi}{N} \cos \left( \frac{\Delta \phi}{N}\right) \phi(t) \right] = 2\gamma \partial_t \phi_{in}(t),
    \label{Eq:EOM}
\end{equation}
where the mode resonant frequency is given by 
\begin{equation}
    \omega_0^2 = \omega_L^2 + \frac{\omega_J^2}{N} \cos\left(\frac{\Delta \phi}{N}\right) = \frac{1}{N C} \left[ \frac{1}{L_s} + \frac{1}{L_J} \cos \left(\frac{\Delta \phi}{N}\right) \right].
\end{equation}


\begin{figure*}[h!]
\centering
\includegraphics[width = 6.5in]{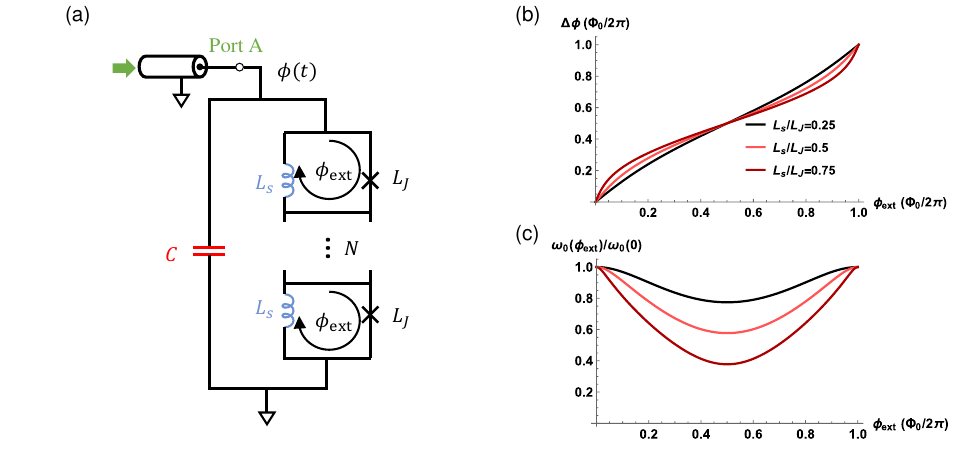}
\caption{
(a) The design with labeled node phases and circuit parameters under consideration for periodic steady-state solutions of the equation of motion. (b) Illustration of the relation between external flux bias to the total internal DC flux through the rf-SQUID ring. Amplifiers that have a higher $\beta = L_s/L_j$ ratio are more nonlinear. (c) The nonlinearity manifests in the tunability of the amplifier, with higher shunting ratios showing more flux modulation.
}\label{appfig:TheoryDesign}
\end{figure*}


%


To calculate the pump power that produces $20\dB{}$ gain, we apply a procedure similar to the tuneup of the real amplifier. At a particular flux value, we can turn up the pump power and analyze the gain of a weak signal. Repeating this process over a half-quanta of flux values produces a plot shown in Fig.~\ref{appfig:TheoryConfigFig}(a), showing a contour of gain points that are all viable bias points. We have divided this contour into two pieces, representing the lower pump power $20\dB{}$ gain points in green, and the higher pump power points in red. Despite also looking like $20\dB{}$ gain points, points along the red contour are not guaranteed stability and so we neglect them in the analysis, continuing with only the bias points along the green curve. To investigate the variation over the flux bias of the simulated amplifier, we move along the green contour, testing both the saturation power $P_{-1\text{dB}}$ and the core power added efficiency

\begin{equation}
\eta_{PAE} = \frac{P_{\text{out},1\text{dB}}(\omega_s)-P_{\text{in}}(\omega_s)}{P_{\text{pump}}(\omega_p)},
\label{Eq:PAE}
\end{equation}
(which reduces to approximately $P_{\text{out},-1\text{dB}}/P_{\text{pump}}$ at $20\dB{}$ gain and near amplifier saturation) at each bias point of the amplifier, with the results shown in Fig. \ref{appfig:TheoryConfigFig}(b)-(c). We find only weak variation in saturation power, but more pronounced variation in the pump efficiency at saturation.

The marginal variation is seen along the bias flux for saturation power, reinforcing notions put forward in \cite{Sivak2019} and \cite{Liu2020} that the static Kerr of the amplifier mode does not act as the primary control of the saturation power, however the significant variation in pump power without a change in saturation power also complicates attributing saturation solely to pump depletion. It may be possible, for example, that the effects of Kerr and pump depletion are opposed to one another in how they affect saturation, with the low (static) Kerr bias points toward $\phi_\text{ext}\approx 0.25\phi_0$ having more saturation power relative to the pump, but less pump power to draw from at $20\dB{}$ gain. Continued study of this equation of motion is necessary to explain these effects in greater detail~\cite{Hougland2024}.

\begin{figure*}[h!]
\centering
\includegraphics[width = 6in]{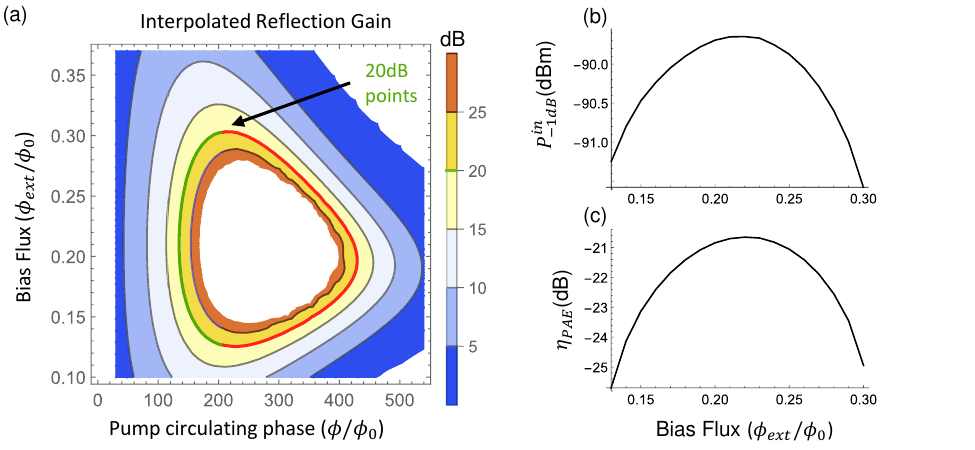}
\caption{
(a) The biasing space of a given parametric amplifier, in this case matching the simulation parameters of the amplifier in the main text with a shunt inductance of $15\pH{}$, a junction inductance of $60\pH{}$, and a quality factor of $8.1$, with 25 rf SQUIDs arrayed in series. The green highlighted contour indicates the powers that are selected for the rest of the analysis in (b) and (c). 
(b) Extracted $P_{1dB}$ along the green contour shown in (a).  (c) Variation in pump efficiency with respect to the bias flux.}
\label{appfig:TheoryConfigFig}
\end{figure*}
\begin{figure*}[h!]
\centering
\includegraphics[width = 6in]{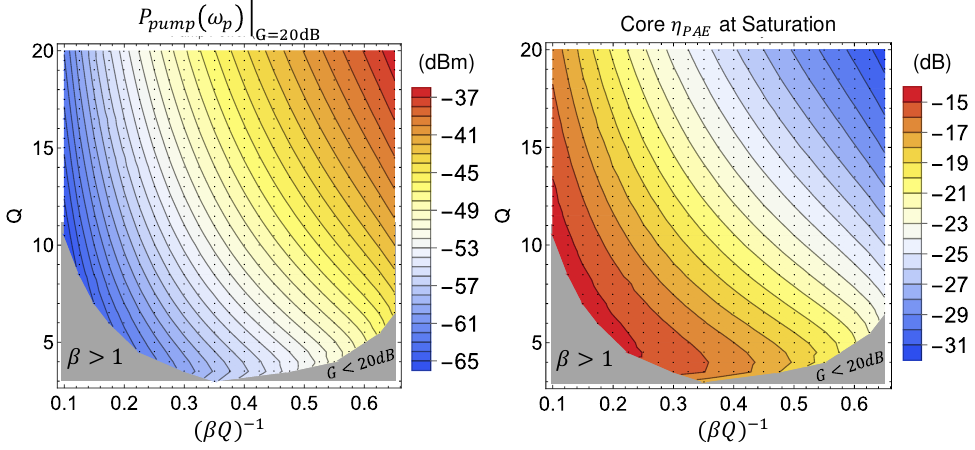}
\caption{
(a) The simulated required pump power of the design space of the JPA's in this work. (b) A similar diagram to the previous one for pump efficiency.
}\label{appfig:pumpEfficiencyTheory}
\end{figure*}
Lastly, we sweep through a significant subset of all possible families of amplifiers of this circuit type to investigate how both the pump power added efficiency and the pump power required to achieve $20\dB{}$ gain varies. In Fig.\ref{appfig:pumpEfficiencyTheory}(a) we see that within a given gain family of amplifiers (a vertical line cut through either (a) or (b)), the required pump power increases with respect to the quality factor, and in general that the higher saturation power families of amplifiers determined by the x-axis coordinate require disproportionately higher pump powers at higher quality factors.  

We can see this disproportionality directly in Fig.\ref{appfig:pumpEfficiencyTheory}(b), which shows that the pump efficiency drops significantly with higher quality factor and decreases yet more as one designs for larger and larger (higher and higher saturation power) $(\beta Q)^{-1}$. We note that despite the simulated pump efficiency of approximately $-20\dB{}$ for the parameters of our devices, the amplifier in this work performs worse in total efficiency, only achieving approximately $-40\dB{}$ pump efficiency because of the pump reflecting off of the coupling capacitor, discussed in Appendix~\ref{"designAppendix"}.

    \section{Pump efficiency}\label{"designAppendix"}

The advantage of utilizing a 3-wave JPA over a 4-wave JPA is the ability to use a single pump at twice the signal frequency, generally reducing the pump leakage into the signal band of the readout chain and making it easier to implement filters to further reduce pump leakage. However, this advantage comes at a cost, which is that the behavior of the circuit near the pumping frequencies has to be carefully considered independently of the behavior at the signal frequencies. For example, it is possible that far away from the signal band it is challenging to get the pump tone to interact with the parametric core of the system that you are driving, significantly raising the pump power requirements of the amplifier and possibly heating the dilution refrigerator it resides in. Therefore, it is important to consider how effectively a certain input pump power drives parametric processes. 

We have already simulated $\eta_{PAE}$ (shown in Eq.~\ref{Eq:PAE}) for the core of the amplifier without the capacitive coupler, shown in Fig.~\ref{appfig:pumpEfficiencyTheory}(b), and so we will define another quantity that captures the difference between the input coupling at the signal frequency, captured effectively by the decay rate $\gamma$, and the pump frequency. This difference is not modeled in the periodic steady-state simulation, which uses the same decay rate for both signal and pump. To model the real circuit, which uses a coupling capacitor to alter the real part of the impedance on resonance, we will modify the "core efficiency" shown in Fig.~\ref{appfig:pumpEfficiencyTheory}(b) by another factor which we will calculate as a function of any coupling impedance.

As detailed in Appendix~\ref{"theoryAppendix"}, the driving term we are concerned with is the circulating phase at the core node of the amplifier, which is the pump frequency component of the phase $\phi_{in}$ across the rf-SQUID array element. To model the effect of any arbitrary coupling series impedance, we consider the circuit below in Fig.~\ref{appfig:pumpEfficiencyCircuit}.

\begin{figure}[h!]
\includegraphics[width = 2in]{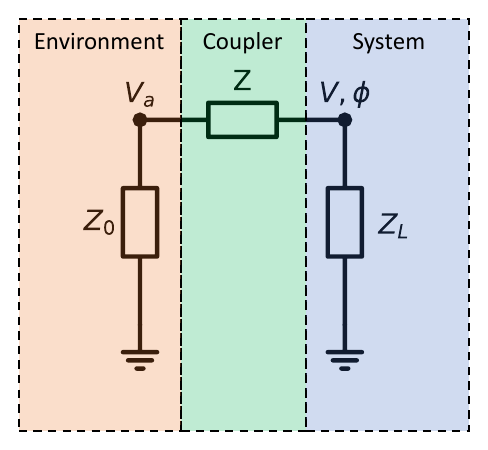}
\caption{
The coupling circuit to consider for evaluating pump power needed outside the circuit, dissipating in a real impedance $Z_0$, versus the circulating phase $\phi$ across the load impedance $Z_L$ as a function of the coupling impedance $Z$. 
}\label{appfig:pumpEfficiencyCircuit}
\end{figure}

For the particular case of a series coupling impedance, the result is easily attainable by voltage division, allowing us to find the power dissipating in the real impedance outside the amplifier, given by $P_a = |V_{a,p}|^2/(2Z_0)$, as a function of the impedances and the required circulating phase

$$P_a = \frac{1}{2Z_0}Re[|\frac{\phi_p\Phi_0}{2\pi} i \omega (1+\frac{Z}{Z_L})|^2],$$

\noindent where $\phi_p$ and $V_a$ are complex amplitudes for $\phi(t) = \phi_p e^{-i\omega t}$, $V_a = V_{a,p} e^{-i\omega t}$ respectively. This can be further simplified by assuming all-reactive loads $Z_L=iX_L$ and couplers $Z=iX$, as is the case for a JPA coupled via a capacitor as long as the pump is not dissipating in the amplifier,

$$P_a = \frac{(|\phi_p| \Phi_0\omega)^2}{2 Z_0 (2\pi)^2} \left(1+\frac{X}{X_L} \right)^2.$$

\noindent We can further group the pre-factors with $\alpha = \frac{(\Phi_0)^2}{2 Z_0 (2\pi)^2}$ and rearrange the equality to form an efficiency metric 

$$\eta_{PCE}:=\frac{\alpha\omega^2|\phi_p|^2}{P_a} = \left(\frac{X_L}{X+X_L}\right)^2.$$

The efficiency that the core utilizes the pump which makes it to the core node is the topic of dedicated research of its own \cite{Hougland2024}, so we define this simpler metric as phase coupling efficiency (PCE). This quantity allows the circuit designer to accurately anticipate the amount of pump power that will make it to the core of the amplifier under some assumptions, and serve as useful energy to the amplification process as opposed to being reflected off of the coupler. The phase coupling efficiency, pump efficiency, and targeted output saturation power combine to determine the total power requirement of the amplifier, the relevant quantity when considering heating problems in a dilution refrigerator. 

\begin{equation}
\eta_{total} = \eta_{PCE} \cdot \eta_{PAE}
\end{equation}

There are some limitations to this simple calculation, the chief among them is that there must not be a zero in the denominator, or else the efficiency diverges positively at that particular frequency. At first, this appears to be a good thing. However, a positive divergence necessarily implies that the pump power to induce \textit{any} phase across the load becomes arbitrarily small. This consequence is incompatible with a stiff pump, and so it appears that the PCE can only be evaluated far away from such poles.

This can be calculated for a coupling capacitor $C_c$ with reactance $X_{Cc} = -1/\omega C_c$ and is most useful in the approximation of driving far above resonance of the mode, where the load reactance of a parallel LC resonator is dominated by the capacitor $C$. This gives a high-frequency limit to the PCE and graphed by the dotted line in Fig.~\ref{appfig:PCE} of
$$\eta_{PCE} = \left(\frac{1}{1+C_c/C}\right)^2.$$

\begin{figure}[h!]
    \centering
    \includegraphics[width=0.5\linewidth]{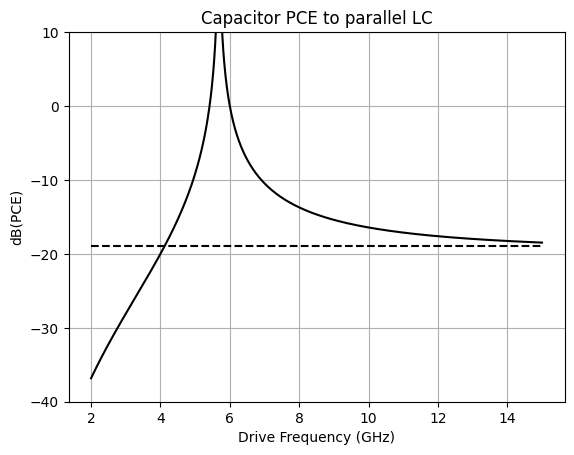}
    \caption{Capacitor PCE for a capacitively coupled JPA with Q = 10. The dotted line indicates the high-frequency limit determined by the ratio of shunt and coupling capacitance. Besides the inherent efficiency of the core, the PCE directly modifies how much external pump power is needed to drive gain for a three-wave JPA, increasing power requirements for 20dB gain by a factor of 10-20dB for the large coupling capacitor configuration in our device.}
    \label{appfig:PCE}
\end{figure}

Power-added efficiency is not usually the dedicated focus of JPA's fabricated in academic settings, with the primary focus usually being on quantum efficiency, bandwidth, and saturation power in that order of priority. Nonetheless, we find it interesting to review the power-added efficiencies of many JPA's in the field. We found the first such review in \cite{Sivak2019} very helpful for perspective on the topic, and so have edited it here with some additions. While we do not make similar distinctions for total versus core power added efficiencies for all devices, due to the many differences in couplers and three versus four-wave nonlinearities employed, we still find it helpful to gauge our device's pump efficiency in reference to other devices fabricated in the field. With this in mind, we conclude that while the core efficiency is competitive at the $\eta_{PAE}=1\%$ level, the choice of pump coupler places the $\eta_{total}$ of our device below the norm.

\begin{figure*}[h!]
\centering
\includegraphics[width = 6.5in]{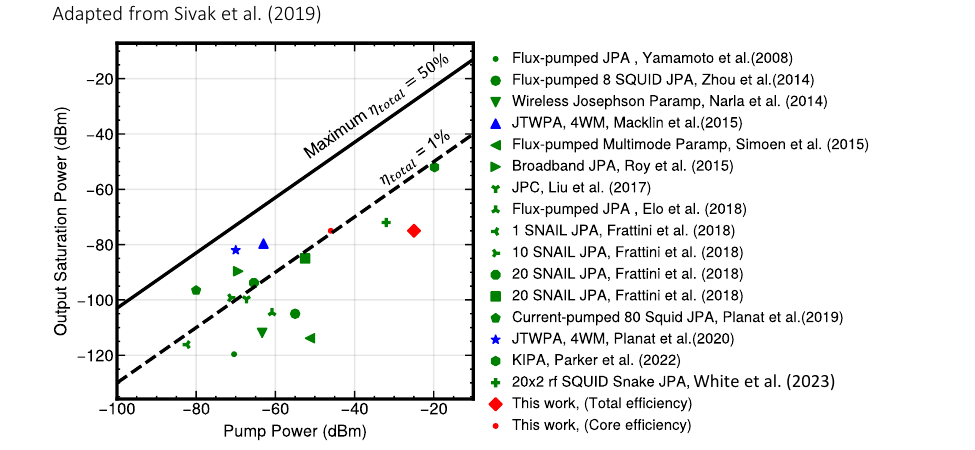}
\caption{
A summary of pump efficiencies of JPAs from other publications, reproduced from \cite{Sivak2020} and expanded. Our devices (red cross) lie well below the trend for other devices as well as our theoretical prediction (red diamond) because of reflections off of the coupling element. 
}\label{appfig:pumpEfficiency_Sivak}
\end{figure*}

The PCE and the PAE determine the total power requirement of the amplifier, and so future work will include coupler optimization to more effectively bring down the power requirement without creating too soft of a pump. Accounting for PCE will hopefully allow future devices to operate without the use of a hot pump line and diplexer.



\section{Experimental details}\label{"exptAppendix"}

\subsection{Fabrication}\label{"sec:fab_layers"}
Our devices were fabricated at Boulder Microfabrication Facility (BMF) at National Institute of Standards and Technology using a Nb/Al-AlOx/Nb trilayer process with low-loss amorphous silicon (a-Si) as an inter-wiring dielectric as described in [25]. From the substrate up the nominal layer thicknesses are as follows: 381-\um{}-thick intrinsic Si wafer (~$20\SI{}{\kilo\ohm\centi\meter}$ resistivity) with 200 nm Nb, 8 nm Al-AlOx, and 110 nm Nb forming a trilayer that is patterned top-down on the substrate to define Josephson junctions and the bottom Nb wiring layer. These structures are blanketed by 310-nm-thick a-Si dielectric layer. Vias patterned through the dielectric form contacts between the junctions, or bottom wiring, and the 300-nm-thick Nb deposited as a top wiring layer over a-Si. The top and bottom wiring layers together with the low-loss a-Si between them form thin-film parallel-plate capacitors. The fabrication was further modified from the description in~\cite{Lecocq2017} to insert a thin bi-layer (< 20 nm) of Nb/Al between a-Si and the top Nb wiring layer to facilitate improved patterning of layers and increase device yields. The full details of the fabrication will be provided elsewhere. The top Nb wiring is bonded onto an external PCB with 1\% Al-Si wirebonds to carry signals off-chip into the external environment.

\subsection{Lines and external circuit elements}\label{"sec:lines"}
From its single input port, the device is driven through two different lines via a frequency diplexer with a low-pass $3\dB{}$  cutoff at $8.65\GHz{}$, a high-pass $3\dB{}$ cutoff at $9.35\GHz{}$, and over $80\dB{}$  rejection from the high-pass port to the common port at the frequencies in this data. The high-pass line provides the pump through a very lightly attenuated ($20\dB{}$ attenuator at $4\K{}$) line, and the low-pass port receives the reflected signal from the cavity via the circulator and two Keycom 15\cm{} superconducting NbTi semi-rigid coax cables. The diplexer enables rejection of the thermal noise of the high-power pump line without losing pump power at the pump frequency at the cost of additional cabling and insertion loss of approximately $1\dB{}$ from the diplexer, both of which must reduce the quantum efficiency of the measurement.

\begin{figure*}[h!]
\centering
\includegraphics[width = 6in]{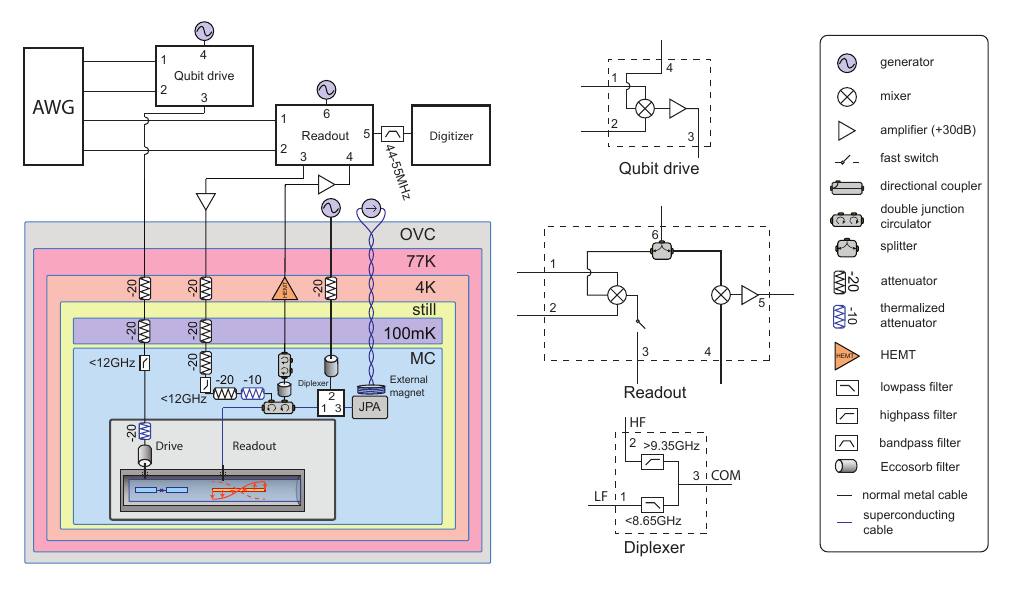}
\caption{
The full circuit diagram of the high-power rf-SQUID JPA experiment.
}\label{appfig:lines}
\end{figure*}

\subsection{Flux biasing: offsets}\label{"sec:flux offsets"}
This amplifier is sensitive to external flux offsets caused by the earth's magnetic field as well as other stray magnetic fields such as nearby ferrite circulators and isolators. In order to shield the device from such stray fields, the amplifier package is enclosed first by a thick 6061 aluminum superconducting shell and then again by a high permeability cryoperm shield made from Amumetal 4K. Despite these precautions, we are able to see distortions in some setups, such as flux offsets even when no bias current is running through the superconducting electromagnet used to bias the amplifier, seen in Fig.~\ref{appfig:fluxsweeps}(b). However, freshly annealed cryoperm cans are able to reject these external flux offsets effectively, shown for a different device in Fig.~\ref{appfig:fluxsweeps}(a) by the high degree of symmetry around zero bias current.

\begin{figure*}[h!]
\centering
\includegraphics[width = 6in]{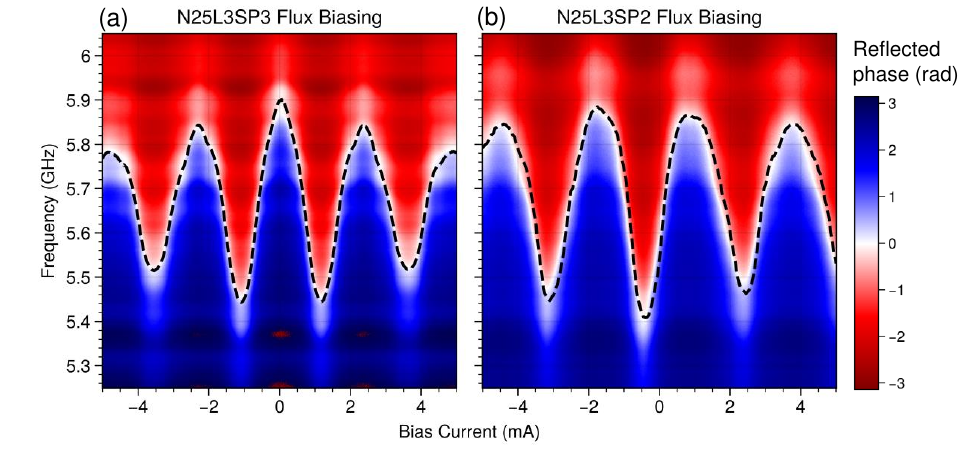}
\caption{
(a) The flux sweep of one of the two tested devices that make up the data in the main text. The first flux sweep behaves as we expect an rf-SQUID should, with the maximum frequency at zero external bias field. The oscillation in frequency damps at higher fluxes, which suggests that the array is coupling to the external field in an asymmetric way and/or that the field is not homogeneous over the length of the array.
(b) The flux sweep of the device tested for quantum efficiency in this work. We find that in this particular device, the zero bias flux is not at the highest frequency of the device consistently over multiple cooldowns, likely due to damage on an older cryoperm can that was used for shielding.
}\label{appfig:fluxsweeps}
\end{figure*}

\subsection{Flux biasing: tunability}\label{"sec:gain tuning"}
One item that weighs on the choice of the quality factor in the design of the amplifier is the tunable frequency range over which the amplifier can achieve $20\dB{}$ gain. Because of the low instantaneous bandwidth of a typical reflection JPA, their chief practical strength lies in their tunability to whichever frequency the readout cavity needs to be driven for measurement. Because the readout cavity frequency of a typical transmon qubit is fixed and designs vary over a few GHz for modern transmons, it is necessary to be able to tune the gain center frequency of the amplifier on the order of a few hundred megahertz or to broaden the amplifier's gain response using an impedance matching circuit to minimize the number of separate amplifiers required to cover the entire readout band. With this in mind, we observe that the amplifiers in the family that barely achieve $20\dB{}$ gain shown on the right-hand side of Fig. 2 only achieve that gain in a narrow range of bias flux. Despite having the best saturation power, these devices would effectively only have a tunable bandwidth comparable to their instantaneous bandwidth, motivating increasing the shunt inductance slightly to favor a more practical tunable bandwidth. Our devices are simulated to achieve gain for an array inductance of approximately $320$ to $420\pH{}$, which gives a nominal tunable range of approximately $700\MHz{}$ measured from the maximum frequency.

In practice, however, the device achieves a much narrower tunable bandwidth than this, only about $150\MHz{}$, which can be seen in Fig.\ref{appfig:satPwrVsBias}(b). One would expect that perhaps the reduction in tunable bandwidth could arise from flux nonuniformity in the array, and so we analyze this possibility next.

\subsection{Flux biasing: uniformity}\label{sec:flux uniformity}

Our high-power JPA is biased by the use of a superconducting electromagnet set into a cutout underneath the copper amplifier package. With a \mm{}-scale distance from the plane of the array to the edge of the electromagnet solenoid and the finite radius of the coil of approximately $1\cm{}$, we anticipate that there should be very little flux nonuniformity in the array from the nonuniformity in the magnet field alone. 

To investigate this, we sweep the current through the magnet over the maximum possible values without causing heating of the magnet loom connections in the refrigerator, then analyze a fast Fourier transform (FFT) of the resonant frequency modulation in Fig.~\ref{appfig:fluxsweeps_FFT}(b). We can then identify peaks in the spectrum in Fig.~\ref{appfig:fluxsweeps_FFT}(b). The transform correctly identifies the primary modulation frequency highlighted by a green vertical line about the peak value. 
\begin{figure*}[h!]
\centering
\includegraphics[width = 6in]{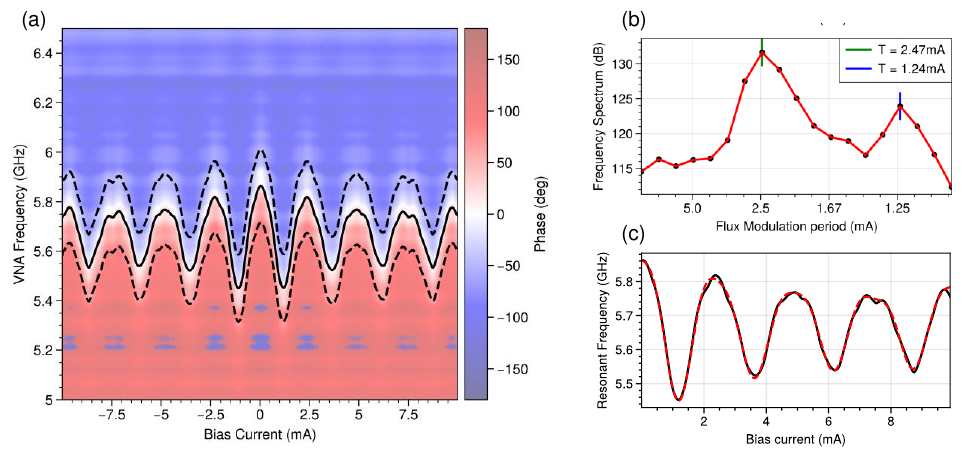}
\caption{
(a) A wider fluxsweep of the device shown in Fig.~\ref{appfig:fluxsweeps}(a) that is suitable for the multiple periods required for accurate analysis in the Fourier domain. Fitted resonant frequency is shown by the solid black line, and the fitted quality factor at each flux bias is indicated by the bracketing dotted lines.
(b) A discrete cosine transform of the fitted resonant frequency, truncated approximately $20\dB{}$ below the highest peak and convolved with a 3-sample wide Blackman window, shown on a log scale. The modulation period of the highest peak is the closest analog of a single flux quanta threading each loop in the array. However, it is clear that not all of the rf SQUIDs are in sync with one another. (c) The resonant frequencies are compared directly with an inverse discrete cosine transform of (b), serving to verify that the smoothing and truncation of the full transform have not appreciably altered the data.
}\label{appfig:fluxsweeps_FFT}
\end{figure*}

The origin of the second-highest peak is unknown. One possible explanation is that large superconducting planes at the boundaries of the array are shielding the first and last few rf SQUIDs from receiving the full external flux through their loops. This local shielding would be very similar on both the ground and resonator node boundaries of the array.  In effort to remedy this, future efforts will include tapers to the array, diminishing the local effect of any adjacent superconductors. 

\subsection{Saturation power calibration}\label{"sec:sat pwr cal"}
Power calibrations are often difficult to do in a straightforward way for devices at the base of a dilution refrigerator. Measurements of the input attenuation up to the plane of the amplifier at room temperature are easy, but they are often inaccurate, as all of the losses are susceptible to changes in temperature. The most straightforward way to calculate power for an amplifier serving to read out a qubit is to use the qubit as a power meter \cite{Schuster2007}. Independent measurements of the qubit can extract the $\chi$ frequency shift of the qubit per photon in the readout cavity. This allows qubit spectroscopy at varying readout powers to directly measure steady-state power in the readout resonator by watching the qubit shift in frequency. As we show in Fig.~\ref{appfig:expt_powerCal}(a), after establishing this baseline with the qubit, the only tasks remaining are to link the DAC drive voltage to a voltage on a spectrum analyzer. This is simple to do by simply removing the drive from the fridge and attaching it to a spectrum analyzer, a necessary step in eliminating mixer LO leakage and tuning single-sideband upconversion in any case. The results of these calibrations as well as similar procedures for the VNA are shown in the remaining subplots in Fig.~\ref{appfig:expt_powerCal}. In our case, the conversion loss of the upconversion mixer and the cable losses are roughly balanced by the additional amplification in the chain. 

\begin{figure*}[h!]
\centering
\includegraphics[width = 6in]{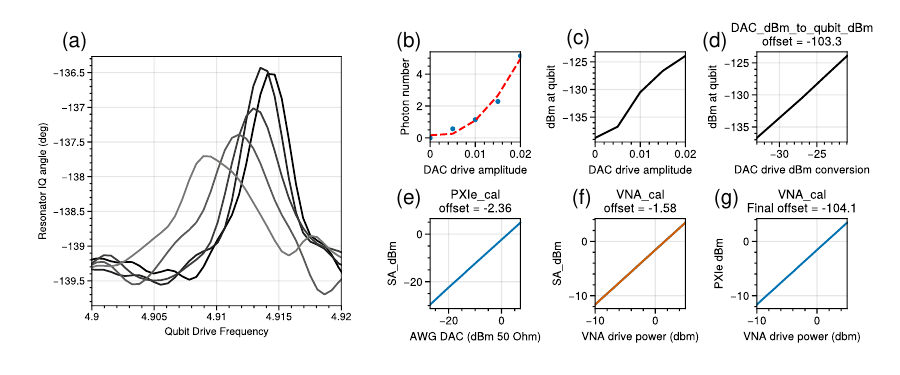}
\caption{
(a) Qubit spectroscopy with varying readout resonator power. The $\chi$ and $\kappa$ of the qubit are independently calibrated using standard reflectometry and measurement post-selection.
(b) Extraction of steady-state resonator photon number versus DAC voltage drive, exhibiting the quadratic variation expected. (c) conversion of the previous plot to dBm (d) Conversion of the x-axis of the previous plot to dBm. This forms the principle part of the calibration (e)-(g) Calibration the rest of the measurement chain against the spectrum analyzer using the same cable that they would connect into the dilution refrigerator with, and subtracting them to amount to a small change to the final offset. Powers are swept over a wide range to check the linearity of the measurement chain. 
}\label{appfig:expt_powerCal}
\end{figure*}

\subsection{Saturation power variation over bias points}\label{"sec:sat pwr variation"}
Many amplification schemes focus on the available bandwidth of the amplifier. Considering resonant JPA's without matching networks typically lack instantaneous bandwidth, they need to be flux tuned to different operating points, and the pump moved in frequency, in order to achieve gain at a variety of frequencies. A figure of merit of such a device is the tunable bandwidth - the range where the amplifier can operate when allowing for any possible tuning. Because changing the flux bias and pump tuning of the amplifier likely affects the saturation power, we measure multiple bias points distributed over the rf-SQUID JPA's tunable bandwidth of approximately $150\MHz{}$.
\begin{figure*}[h!]
\centering
\includegraphics[width = 6in]{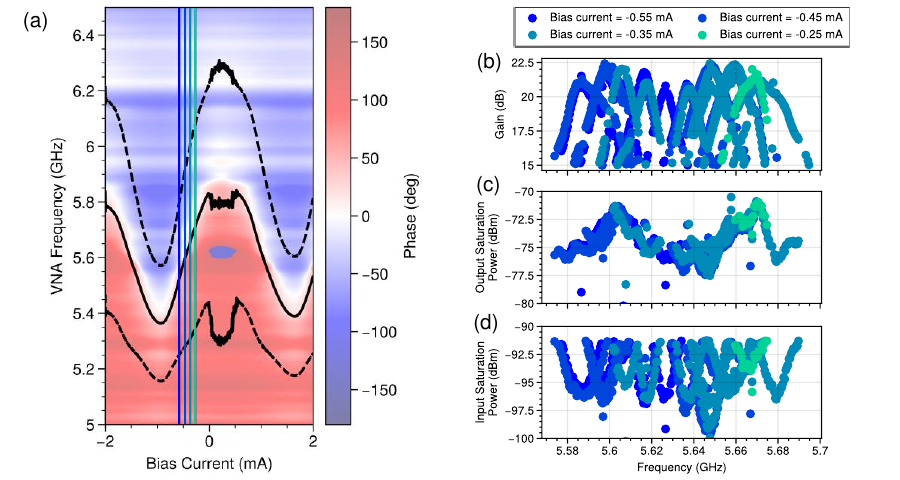}
\caption{
(a) A flux sweep of the device used for saturation power data in this work, with vertical lines marking particular bias points of interest for producing gain in the amplifier.
(b)-(d) Various gain bias points of the amplifier used for saturation data after accounting for the maximum possible loss at the given frequency over any flux bias of the amplifier, shown with their measured input and output saturation powers. The colors of the points correspond to their flux bias points.
}\label{appfig:satPwrVsBias}
\end{figure*}

We find some few-dB variation in the saturation power over the bias points of the amplifier, with most bias points achieving close to the average value reported in the main text. This is supported by the simulations, and similar to many previously reported results for array amplifiers \cite{Frattini2018, Sivak2019, Sivak2020}. 

\subsection{High power measurement transitions}\label{"sec:high pwr transitions"}

A natural next step in qubit measurement with a high-power JPA is to increase the power of the measurement to attain a higher integrated SNR with a shorter measurement time. While the amplifier is more than capable of handling the additional power, up to several hundred photons at this readout resonator frequency, the readout resonator plus qubit system is certainly not, as discussed in the main text. Here we show an example of how this process can go awry by showing integrated readout histograms.

In Fig.~\ref{appfig:highPWRMsmt}(c) and (d) we give just one example of the measurement results one might expect to see when measuring at powers where the QND nature of dispersive measurement breaks down because of measurement-induced transitions in the qubit state. Transitions occurring faster than the integration time of the readout (here approximately 800ns, not including the ring-down time of the resonator) are visible as the ``tracks'' between states, increase in frequency at higher powers, as noted in multiple other publications \cite{Sank2016, Shillito2022, Khezri2022, Cohen2023}. In particular, we even see transitions that are supposed to be forbidden, as well as many multi-photon transitions. In this case, the time resolution to view these transitions is limited by the ring-up and ring-down time of our$~1\MHz{}$ linewidth resonator, and so we leave the study of these effects in further detail to future work.

\begin{figure*}[h!]
\centering
\includegraphics[width = 6in]{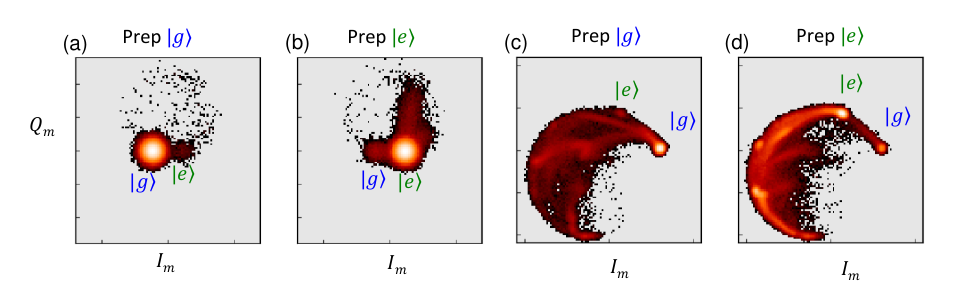}
\caption{
(a) and (b) measurement histograms of the transmon shown in the main text with the amplifier on, counts shown in logarithmic color scale. 
(c) and (d) High power measurements of the transmon in the main text. Because of the high saturation power of the rf-SQUID JPA, we can probe the extremely high power response of the transmon. Multiple higher states and transitions between states are easily separable, enabling future studies of transition rates and effective qubit temperature as a function of readout power.
}\label{appfig:highPWRMsmt}
\end{figure*}

\end{widetext}

\end{document}